\documentclass[10pt]{article}

\usepackage{geometry}
\usepackage[utf8]{inputenc}
\usepackage{graphicx}
\graphicspath{ {images/} }
\usepackage{booktabs} 
\usepackage{array} 
\usepackage{paralist} 
\usepackage{verbatim} 
\usepackage{mathtools}
\usepackage{amsmath,amssymb,amsthm}
\usepackage{mathrsfs}
\usepackage{microtype}
\usepackage{hyperref}
\usepackage{cleveref}
\usepackage{slashed}
\usepackage{booktabs}
\usepackage{array}
\usepackage{bbold} 
\usepackage[justification=centering]{caption}
\usepackage{subcaption}
\usepackage{tikz}
\usepackage{titlesec}
\usepackage{titling}
\usepackage{authblk}

\newcommand{\mc}[1]{\mathcal{#1}}
\newcommand{\ii}{\text{i}}

\newcommand{\dd}{\textrm{d}}
\newcommand{\db}{\bar{\partial}}

\swapnumbers

\newtheorem*{theorem*}{Theorem}

\newtheorem*{prop*}{Proposition}

\newtheorem*{conj*}{Conjecture}

\newtheorem*{prin*}{Principle}

\theoremstyle{definition}

\newtheorem*{defn*}{Definition}

\newtheorem*{warning*}{Warning}

\pretitle{\begin{center}\fontsize{12bp}{10bp}\bfseries}
   \posttitle{\par\end{center}}
\preauthor{\begin{center}\fontsize{10bp}{10bp}\scshape}
  \postauthor{\par\end{center}\vspace{0bp}}

\titleformat{\section}{\centering\bf}{\thesection .}{0.4em}{}
\titleformat{\subsection}{\it}{\thesubsection}{0.4em}{}
\titleformat{\subsubsection}{\it}{}{0em}{}

\title{Exotic vortices and twisted holomorphic maps}
\author{Edward Walton\thanks{\href{mailto:e.walton@damtp.cam.ac.uk}{e.walton@damtp.cam.ac.uk}}}
\affil{\small{\textit{Department of Applied Mathematics and Theoretical Physics,}\\
\textit{University of Cambridge,}\\
\textit{Wilberforce Road, Cambridge, UK}}\\ \vspace{-1cm}}
\date{}

\begin{document}

\maketitle

\begin{abstract}	
We consider the exotic vortex equations on compact Riemann surfaces. These generalise the well-known Jackiw--Pi and Ambj{\o}rn--Olesen vortex equations and arise as equations for Bogomolny--Prasad--Sommerfield-like configurations in nonrelativistic Chern--Simons-matter theories. We show that (exotic) vortex solutions in \(U(N)\) gauge theories with a single fundamental flavour on compact Riemann surfaces can be constructed from gauged holomorphic maps into complex space forms of dimension \(N\). We conjecture, with some evidence, that all such solutions arise this way. This leads us to insights regarding the moduli theory of exotic vortices, including the identification of interesting selection rules for nonAbelian exotic vortices on the sphere. We also consider the classical and quantum dynamics of exotic vortices. Using recent localisation results, we give the Witten index of the quantum mechanics describing low-temperature Abelian exotic vortex dynamics. 
\end{abstract}

\section{Introduction and summary}

Exotic vortices are vortex-like configurations in physical systems which exhibit a kind of `anti-Meissner--Higgs effect'. Rather than acting to screen magnetic fields, the Higgs fields in these theories conspire to accumulate in regions of large magnetic flux. This means that `inside' a vortex, where the Higgs fields become small, the magnetic field is at its weakest.

The equations for BPS-like exotic vortices are exactly as the usual vortex equations, constraining the configuration of a gauge potential \(A\) and a Higgs field \(\phi\) living on a Riemann surface \(\Sigma\), with schematic form
\begin{equation}
\label{eq:ev}
\begin{gathered}
*F(A) = \frac{\ii}{2} e^2(\tau - \phi \phi^\dagger)  \\
\db_A \phi = 0 \text{,}
\end{gathered}
\end{equation}
except that the constant \(e^2\), which we usually think of as a gauge coupling constant, is allowed to be nonpositive (we give a more complete description below). 

Integrating the trace of the first vortex equation over a compact surface tells us that vortices `take up' some area proportional to \(e^2\). For conventional vortices, with \(e^2 > 0\), this leads to the Bradlow bound, constraining the number of vortices that can fit on a compact surface of finite area. When \(e^2 <0\), the vortices `produce' area: there is no upper bound on the number of vortices.

As we summarise below, exotic vortices arise most naturally in nonrelativistic Chern--Simons-matter theories. Abelian examples include the Jackiw--Pi vortex equation \cite{jackiwSSS} and the equations studied by Popov \cite{popovIV} and Manton \cite{mantonPV,mantonFVE}. Among other equations, including that of Jackiw--Pi, Manton studied the so-called Ambj\o rn--Olesen equation, which originally appeared in descriptions of the electroweak phase transition \cite{ambjornAS,ambjornMCS,ambjornCS}. Abelian exotic vortices in the presence of magnetic impurities have been studied in \cite{gudnasonMI}.

While Abelian examples are most prominent in the literature, exotic vortices can be considered for the data of a general gauge theory \cite{dunneSCS,turnerQO} and we recall the basic theory here, extending it to allow for matter living in general nonlinear symplectic representations of the gauge group.

We will address the problems of finding solutions to these equations, understanding the moduli spaces of solutions, and understanding the (quantum) dynamics of exotic vortices. In doing this we address some of the questions raised in \cite{doroudSA} in the context of Jackiw--Pi vortices. 

Our main technical result, which is based on results of Manton \cite{mantonFVE}, Witten \cite{wittenMPS}, and Olesen \cite{olesenSC}, relates solutions to the (exotic) vortex equations to objects that we call \emph{twisted holomorphic maps} into complex space forms (complete K\"ahler manifolds of constant holomorphic sectional curvature). For us, a twisted holomorphic map from a Riemann surface \(\Sigma\) into a K\"ahler manifold \(X\) is a flat \(\text{Isom}(X)\)-bundle equipped with a holomorphic section of the associated holomorphic \(X\)-bundle. We explicitly construct a map from the space of equivalence classes of twisted holomorphic maps from \(\Sigma\) to \(X\) to the space of equivalence classes of solutions to the exotic vortex equations \eqref{eq:ev} on \(\Sigma\) with gauge group \(U(\text{dim}_\mathbb{C}(X))\), one fundamental flavour, \(e^2 = -\kappa\), and \(\tau = \frac{2-2g}{2\pi \text{vol}(\Sigma) \kappa} \mathbb{1}\). 

We further conjecture, with both local and global evidence, that the converse holds: every solution to the relevant vortex equation arises in this way and the moduli space of the vortex solutions is in one-to-one correspondence with the moduli space of relevant twisted holomorphic maps. As the moduli space of twisted holomorphic maps is rather more accessible that that of exotic vortices, this conjecture allows gives us conjectural information about the moduli space, and therefore the quantum mechanics, of exotic vortices. 
In particular, our results give additional context to results of Turner on nonnormalisable modes in the quantum mechanics of Jackiw--Pi vortices \cite{turnerQO}. The conjecture also implies certain selection rules for exotic vortices on the sphere, generalising the known fact that Jackiw--Pi vortices on the sphere have even vortex number \cite{jackiwSSS}. In our context, such results are related to the topology of flat \(\text{Isom}(X)\)-bundles.

Our results have something to say about conventional (nonexotic) vortices in \(U(N_c)\) gauge theories with a single fundamental flavour, generalising Witten's work \cite{wittenMPS} for \(N_c = 1\). We focus our attention primarily on the exotic case of \(e^2<0\), as, in this case, the twisted holomorphic map moduli problem is more accessible. 

The paper continues as follows. In the rest of the introduction, we summarise the story in more detail, outlining our main results and pointing out some related work. In \autoref{sec:evm}, we describe the exotic vortex equations in detail and give some basic facts about the structure of their moduli spaces of solutions. In \autoref{sec:thc}, we prove our main result and use it to give conjectural descriptions of exotic vortex moduli spaces. We show how previous results, such as Olesen's single-vortex solution to the Jackiw--Pi equation on the torus \cite{olesenSC}, fit into our work. In \autoref{sec:dyn}, we describe the (quantum) dynamics of gauge theories supporting exotic vortices. The natural classical dynamics is first order. At low temperature, it is described by Hamiltonian mechanics on the moduli space of vortices. Using results of \cite{waltonVC}, we (formally) compute the `expected' (or `supersymmetric') partition function (that is, the Witten index) of the dynamical exotic Abelian theory on Riemann surfaces of arbitrary genus.

\subsection{Exotic vortices from twisted holomorphic curves}

Mathematically, the equations \eqref{eq:ev} with \(e^2 <0\) are harder to analyse than the usual vortex equations, which partially reflects the fact that they do not appear to be stable in relativistic physical theories. In fact, the natural relativistic theories that one would write down which exhibit the kind of anti-Meissner effect that is necessary are completely sick, with Euclidean energy functionals unbounded from below. The unboundedness of the relevant functionals from below makes the usual functional analysis techniques hard, or impossible, to implement.

While this sounds unpleasant, there is beauty in the equations. It was noticed by Manton \cite{mantonFVE} that, on constant curvature backgrounds, special cases of the exotic vortex equations in theories with gauge group \(U(1)\) and one fundamental scalar flavour could be solved by meromorphic functions. This is the exotic analogue of Witten's solution of the hyperbolic vortex equations by Blaschke functions \cite{wittenMPS} and it generalises the fact that the Jackiw--Pi equations on the plane are equivalent to the Liouville equation, which is solved by meromorphic functions \cite{jackiwSSS,horvathyVS}. It was shown in \cite{contattoMF} that these results could be understood in terms of dimensional reduction of the self-duality equations in four dimensions, in analogy to Witten's argument in \cite{wittenMPS}. 

We generalise Manton's result in several ways. First, we consider the same vortex systems as Manton, showing that one can produce additional solutions by considering meromorphic functions `twisted' by a flat gauge field for the \(PSU(2)\) symmetry that rotates the target sphere (we explain what this means in detail below). In particular, this includes Olesen's single vortex solution to the Jackiw--Pi equation on the torus \cite{olesenSC}. We conjecture that all of the relevant exotic vortex solutions can be found in this way, giving some coarse evidence for this in the form of an equality of indices. This would be useful, because the moduli problem for a flat \(PSU(2)\) gauge field and twisted meromorphic function is qualitatively easier to understand than that of the exotic vortex system. 

We then generalise the story to nonAbelian theories. We show that, on backgrounds of constant curvature, solutions to exotic vortex equations in \(U(N_c)\) theories with \(N_f = 1\) fundamental flavour can be produced from twisted holomorphic maps into \(\mathbb{C}P^{N_c}\), `twisted' by a flat \(PSU(N_c+1)\) connection. More generally, (not necessarily exotic) vortex solutions can be produced from `twisted' holomorphic maps into K\"ahler manifolds of constant holomorphic section curvature. Usually one asks that \(N_f \geq N_c\) in order to see (nonexotic) vortices, as this is necessary to completely Higgs the gauge group. In the exotic case things are rather different, and we expect to find families of solutions even for \(N_f < N_c\). 

Indeed, we show that the moduli space of pairs consisting of a flat \(PSU(N_c+1)\) connection and a twisted holomorphic curve in the associated \(\mathbb{C}P^{N_c}\)-bundle has the same expected dimension as that of exotic vortices in the relevant \(U(N_c)\) gauge theory with 1 fundamental flavour. This leads us to conjecture that all such exotic vortex solutions on constant curvature backgrounds arise in this way.

We also show how to alleviate the constraint of constant curvature, using `uniformising functions'. While this is formally useful, it takes us a little way from the original simple solution generating techniques of \cite{mantonFVE}.

This construction is novel, but it does not represent the first time that nonAbelian connections have played a role in the story of Abelian exotic vortices. Indeed, the dimensional reduction picture of \cite{wittenMPS,contattoMF} mentioned above involves nonAbelian gauge theory in four dimensions (with gauge group a real form of \(SL(2,\mathbb{C})\) depending on the form of the vortex equation). In \cite{rossMZM}, solutions to the Abelian exotic vortex equations on the two-dimensional sphere (known as Popov vortices, after \cite{popovIV}) were related to flat \(SU(2)\) gauge potentials on the three-sphere. 

A unifying picture might be provided by Hitchin's work \cite{hitchinHM} relating solutions to a kind of `exotic Hitchin system' of equations for an \(SU(2)\) gauge theory to (twisted) harmonic maps into the three-sphere. Our result in the Abelian case can be thought of as arising from this when the \(SU(2)\) connection is reducible to \(U(1)\).

\subsection{Exotic selection rules}

One of the strange things about Jackiw--Pi vortices on the infinite sphere is that there are no solutions with odd flux: the total vortex charge is always even \cite{horvathyVS}. This has no straightforward topological explanation and if one studies the same equation on the flat torus, one can find solutions with odd flux. Such a solution was found by Olesen in \cite{olesenSC}.

While this might look mysterious from the perspective of the vortex equations, it can be understood clearly in the context of twisted meromorphic functions. For us, a twisted meromorphic function is a section \(\psi\) of a \(\mathbb{C}P^1\)-bundle \(S\) over our base Riemann surface \(\Sigma\). The vortex number associated to the section \(\psi\) is 
\[
k = (2g-2) + \text{deg}(\psi^*T_S) \text{.}
\]
This is a kind of generalised Riemann--Hurwitz relation.

If the bundle \(S\) is topologically trivial, then the second term on the right hand side is always even, because \(\text{deg} (T_{\mathbb{C}P^1}) =2\). Then \(k\) is always even. When \(S\) is nontrivial, the second term on the right hand side may be odd.

In general, topological \(S^2\)-bundles on a surface \(\Sigma\) are classified by a Stiefel--Whitney-type class in
\[
H^1(\Sigma, \mathbb{Z}_2)\text{.}
\]
The point here is that the group of oriented diffeomorphisms of the sphere is homotopic to \(PSU(2)\). A \(PSU(2)\)-bundle is locally an \(SU(2)\)-bundle, but one has an additional choice of sign around each noncontractible cycle, giving rise to an element of \(\mathbb{Z}_2\) for each such cycle.

When \(\Sigma = \mathbb{C}P^1\), there are no topologically nontrivial bundles, so \(k\) must be even. When \(\Sigma\) has higher genus, topologically nontrivial bundles are available and so solutions with odd vortex number may exist.

In \autoref{subsec:exnonab}, we show that this generalises to the nonAbelian case of a \(U(N_c)\) theory with one flavour. Now the selection rules are determined by the topology of \(PSU(N_c+1)\)-bundles. When \(\Sigma = \mathbb{C}P^1\), solutions arising from the relevant analogue of the twisted meromorphic functions above are forced to have
\[
k = -2N_c + (N_c+1)p
\]
for \(p \in \mathbb{Z}\). Nontrivial \(PSU(N_c+1)\)-bundles, which exist on surfaces with \(g>0\), allow for \(k - N_c(2g-2)\) to take nonzero values modulo \(N_c+1\).

\subsection{Quantum exotic vortices or: Probing the moduli space with quantum mechanics}

While lots of things that we are used to when thinking about vortices break when one takes \(e^2 < 0\), the construction of the moduli space as a symplectic quotient does not (although it is no longer a K\"ahler quotient). This means that the moduli space should generically be a symplectic manifold, which makes it a good home for Hamiltonian mechanics.

From the point of view of \((2+1)\)-dimensional physics, Hamiltonian mechanics on the vortex moduli space is realised as a low-temperature limit of certain nonrelativistic Chern--Simons-matter theories. This is studied in detail in \cite{waltonVC}. Formally, the calculations go through in the same way, with certain parameters changing sign. 

Because we know very little about the existence of solutions in the general case, these calculations are fundamentally formal and so are highly speculative. However, they are interesting. We compute, for example, the expected degeneracy of quantum Jackiw--Pi vortices at low temperature.

\subsection{Vortices as degenerate metrics}

In \cite{baptistaVDM}, Baptista observed that the usual Abelian vortex equation could be viewed as an equation for a degenerate metric obeying a certain curvature condition. This idea was carried forward by Manton in \cite{mantonFVE}, who noted that solutions to Abelian exotic vortex equations produce metrics with conical singularities with cone angle a multiple of \(2\pi\). 

We will not develop this idea in any detail here, only noting that our results can be used to understand the moduli spaces of constant curvature metrics on Riemann surfaces with conical singularities of the given type (see also \cite{chenCM,eremenkoMC,mazzeoCM}).

\section{Exotic vortices and their moduli}
\label{sec:evm}

\subsection{The exotic vortex equations}

The exotic vortex equations take the same data as the usual, nonexotic, vortex equations. We fix a K\"ahler manifold \(Y\) carrying the Hamiltonian action of a compact Lie group \(G\). We write
\[
\mu : Y \to \mathfrak{g}^\vee
\]
for the corresponding moment map, where \(\mathfrak{g}\) is the Lie algebra of \(G\).

We now take a smooth two-dimensional surface \(\Sigma\) with volume form \(\omega_\Sigma\). We fix a principal \(G\)-bundle 
\[
P \to \Sigma
\]
and form the associated \(Y\)-bundle
\[
\mc{Y} \coloneqq Y \times_G P \to \Sigma \text{.}
\]

We take a section \(\phi\) of \(\mc{Y}\) and a connection \(A\) on \(P\). For a given element \(\xi\) in the dual of the centre of \(\mathfrak{g}\), the vortex equations for \((A, \phi)\) are
\begin{align*}
*F(A) &= e^2(\xi - \mu(\phi)) \\
\db_A \phi &=0	
\end{align*}
where \(e^2\) is a scale factor on the Killing form on \(\mathfrak{g}\).

The \emph{exotic vortex equations} are these equations with \(e^2 \leq 0\). This subtle change from the usual case has dramatic consequences, as we will see. 

The case of \(e^2=0\) is rather trivial, with the equations becoming the equations for a flat connection and a section of an associated holomorphic bundle. Vortices in these theories are sometimes called \emph{Bradlow vortices}, because they resemble vortices `at the Bradlow bound'. They have been considered in \cite{gudnasonSE}. Because they are rather simple from our perspective, we will focus our attention primarily on the case of \(e^2 < 0\).

\subsection{The moduli space as a symplectic quotient}

An important fact about the moduli space of nonexotic vortices is that it is a symplectic, and in fact K\"ahler, quotient (see \cite{garciapradaDEP}). Some mathematical insight into the nature of exotic vortices comes by considering how the quotient construction of the moduli space differs (or rather, how it does not) when \(e^2 \leq 0\). 

In a sense, setting \(e^2 \leq 0\) is all there is to it. Write \(\mc{C}\) for the space of (reasonable) pairs \((A, \phi)\). Recall that this space has a symplectic form, \(\omega_\mc{C}\), defined by
\[
\omega_\mc{C} ((\dot{A}_1, \dot{\phi}_1), (\dot{A}_2, \dot{\phi}_2)) = \int_\Sigma \left( \text{tr} \left( \dot{A}_1 \wedge \dot{A}_2 \right) + e^2 \omega_\mc{Y} (\dot{\phi}_1, \dot{\phi}_2) \omega_\Sigma  \right)\text{.}
\]
There is nothing stopping us from taking \(e^2 \leq 0\) and this is what we do.

This does have consequences, though. When \(e^2 > 0\), the symplectic form \(\omega_\mc{C}\) forms part of a K\"ahler structure. The corresponding Riemannian metric can then be interpreted as a relativistic kinetic energy functional. When \(e^2 \leq 0\), this `metric' is, at best, semiRiemannian (see \cite{mantonFVE,contattoMF}). It may be rather more degenerate than that. A metric with negative norm states can never make for a good kinetic energy functional, because the system will run away to negative energy.

We do not need to worry about this though: we have a symplectic structure on \(\mc{C}\) and the story of gauged Hamiltonian mechanics goes through as in the non-exotic case. The group \(\mc{G}\) of gauge transformations acts symplectically on \(\mc{C}\). Restricting to the space \(\mc{C}_0\) of solutions to \(\db_A \phi = 0\), we take the symplectic quotient to get the vortex moduli space
\[
\mc{M} = \mc{C}_0//\mc{G} \text{.}
\]

In particular, this implies that the exotic vortex moduli space is `generically' symplectic. However, as we will see explicitly, the exotic vortex equations are much less well behaved than the usual vortex equations and simple examples may fail to be generic in this sense.

\section{Exotic vortices from twisted holomorphic curves}
\label{sec:thc}

\subsection{The Abelian case}

In \cite{mantonPV,mantonFVE}, Manton found beautiful integrability results for Abelian exotic vortex equations in the case that the Riemannian surface \(\Sigma\) has constant scalar curvature \(\kappa_0\), building on \cite{wittenMPS,popovIV}. In that case, he produced solutions to the equations 
\begin{align*}
*F(A) &= -\kappa\left(\frac{\kappa_0}{\kappa} - |\phi|^2\right) \\ 	\db_A \phi & =0 \text{,}
\end{align*}
where \(A\) is a \(U(1)\) connection, \(\phi\) is a single scalar field of unit charge, and \(\kappa\) is a positive real number. He did this, roughly speaking, by setting \(|\phi|^2 = *\psi^*\omega_{\mathbb{C}P^1}\), where \(\psi : \Sigma \to \mathbb{C}P^1\) is holomorphic and \(\omega_{\mathbb{C}P^1}\) is the Fubini--Study form, and solving for \(\phi\) in some gauge, showing that the result solved the vortex equations in a second-order form (which is obtained by eliminating the gauge field locally).

We start by generalising the ideas of \cite{mantonFVE} to find more general solutions. To do this, we consider sections of possibly nontrivial \(\mathbb{C}P^1\)-bundles over \(\Sigma\). Indeed, suppose \(S \to \Sigma\) is a \(\mathbb{C}P^1\)-bundle carrying a flat \(PSU(2)\) connection \(a\). Let \(\psi\) be a section of \(S\) obeying the equation
\[
\db_a \psi = 0 \text{.}
\]
The claim is that one can build an Abelian exotic vortex solution from this data. We recover Manton's result when the connection \(a\) is trivial.

We will take a slightly different approach to that of \cite{mantonFVE}. We will work in the `holomorphic gauge', as in \cite{bradlowVLB}, where one views the Abelian vortex equation as an equation for a holomorphic line bundle \((L,\db_L)\) with section \(\phi\) and Hermitian metric \(h\) obeying
\begin{equation}
\label{eq:holgauge}
\begin{gathered}
	*F(h) = \frac{\ii e^2}{2}(\tau - |\phi|_h^2) \\
	\db_L \phi = 0\text{,}
	\end{gathered}
\end{equation}
where \(F(h)\) is the curvature of the Chern connection uniquely associated to \((\db_L , h)\). One can get back and forth between this and the usual vortex equation by the complex gauge transformation \(g\) that sets \(g(h) = 1\).

By the holomorphicity of \(\psi\) and the flatness of \(a\), we have that
\[
\dd_a \psi \in \Omega^0(\Sigma, K_\Sigma \otimes \psi^*T_S)
\]
is holomorphic with respect to the natural holomorphic structure, induced by \(a\) and the holomorphic structure of \(\Sigma\). We set \(L \coloneqq K_\Sigma \otimes \psi^*T_S\).

The round K\"ahler structure \((g_{S^2}, \omega_{S^2})\) defines a vertical K\"ahler structure on the fibres of \(S\). The connection \(a\) allows us to extend the domain of \(g_{S^2}\) and \(\omega_{S^2}\) to include horzontal vector fields, and we write \(g_S\) and \(\omega_S\) for the resulting tensor fields. The flatness of \(a\) means that it can be trivialised locally, so the distinction between \((g_S, \omega_S)\) and \((g_{S^2}, \omega_{S^2})\) is locally insignificant. Note that, despite the notation, \(g_S\) is not a metric on the total space of \(S\) and \(\omega_S\) is not a symplectic form on the total space of \(S\).

The combination of the round Hermitian structure on the fibres of \(S\) and the metric on \(\Sigma\) defines a Hermitian structure \(h\) on \(L\). With this structure, one has
\[
| \dd_a \psi |_h^2 = (g_{S})_{ij} (g_\Sigma^{-1})^{\mu \nu} D_\mu \psi^i D_\nu \psi^j
\]
where we have written \(D \coloneqq \dd_a\) to accommodate the indices. This admits a Bogomolny rearrangement
\[
|\dd_a \psi |_h^2 = 2 |\db_a \psi |^2 + * \psi^*\omega_S
\]
where we have used that \(a\) is flat (if \(a\) is not flat, an extra curvature dependent term appears). Here \(*\) is the Hodge star on \(\Sigma\). Now, the fact that \(\db_a \psi = 0\) tells us that
\begin{equation}
\label{eq:psi1}
|\dd_a \psi|_h^2 = * \psi^*\omega_S\text{.}
\end{equation}

On the other hand, associated to the Hermitian structure \(h\) and the holomorphic structure on \(L\) is the uniquely defined Chern connection \(\nabla_h\). By definition, this has the properties that \(\nabla_h^{0,1} = \db_L\), and \(\nabla_h\) preserves \(h\). The curvature of \(\nabla_h\) is
\[
F(h) = - F(g_\Sigma) + \psi^*F(g_S) \text{,}
\]
where by \(F(g_S)\) we mean the fibrewise curvature of \(g_S\) extended to horizontal vectors using the flat connection \(a\).

Now, \(g_S\) is fibrewise the round metric, so in particular it has constant curvature \(\kappa\), so that \(F(g_S) = \ii \kappa \omega_S\). Combining this with \eqref{eq:psi1} tells us that
\[
F(h) = - F(g_\Sigma) + \ii \kappa * | \dd_a \psi|_h^2 \text{.}
\]
Defining \(\phi\) to be \(\dd_a \psi\), which is a holomorphic section of \(L\), this is starting to look like a vortex equation. The only thing left to do is to ask \(g_\Sigma\) to have constant curvature, so that
\[
F(g_\Sigma) = \ii \kappa_0 \omega_\Sigma \text{.}
\]
As \(F(g_\Sigma)\) is the curvature of a connection on \(T\Sigma\), the Gauss--Bonnet theorem means that we must have
\[
\kappa_0 = \frac{2-2g}{2\pi \text{vol}(\Sigma)}
\]
where \(g\) is the genus of \(\Sigma\).

In this case, we have
\begin{align*}
F(h) &= - \ii \kappa \left( \frac{\kappa_0}{\kappa} -  | \phi|_h^2 \right) \omega_\Sigma \\
\db_L \phi	&= 0 \text{,} 
\end{align*}
where \(\kappa\), being the curvature of a round sphere, is necessarily positive. These are exotic vortex equations in the holomorphic gauge \eqref{eq:holgauge}. Thus \((\db_L, h, \phi)\) defined this way give solutions to the exotic vortex equations. We can put them into the unitary form by acting with a complex gauge transformation \(g: \Sigma \to \mathbb{C}^*\) with \(gg^* = \frac{1}{h}\). 

The vortex number \(k\) is the degree of the bundle \(L = K_\Sigma \otimes \psi^*T_S\). This is given by a generalised Riemann--Hurwitz formula, which we can derive by considering the exact sequences
\begin{align*}
0 \to \mc{O}_S \to p^*E &\otimes \mc{O}_S(1) \to T_{S/\Sigma} \to 0\text{,} \\
0 \to T_{S/\Sigma} \to &T_S \to p^*T_\Sigma \to 0
\end{align*}
of holomorphic vector bundles, where \(p : S \to \Sigma\) is the projection, \(E\) is a choice of rank 2 vector bundle such that \(S\) is the projectivisation of \(E\) (which always exists for dimensional reasons), and \(\mc{O}_S(1)\) is the corresponding tautological bundle.
The bundle \(E\) is not uniquely specified: it can be shifted by any line bundle. This is accompanied by a corresponding shift of the tautological bundle, so that \(p^*E \otimes \mc{O}_S(1)\) is uniquely specified by \(S\), as it must be. The first sequence is the relative Euler sequence, and the second is the Atiyah sequence.

Using these sequences and the multiplicativity of the Chern character, we have that
\[
\text{ch}(\psi^* T_S) = \text{ch}(T_\Sigma) \text{ch}(E \otimes \psi^* \mc{O}_S(1)) 
\]
where we have pulled the sequences back along \(\psi\). From this we can read off the degree of \(\psi^*T_S\). Without loss of generality, we can assume that \(\text{deg}(E) \) is either 0 or 1 (twisting with a line bundle allows us to bring it into one of these forms). If \(E\) has degree 0, this gives
\[
\text{deg}(\psi^*T_S) = 2 \text{deg}(\psi) + 4 - 4g
\]
so that the vortex number is
\begin{align*}
k &= \text{deg}(K_\Sigma \otimes \psi^*T_S) \\
	&= 2\text{deg}(\psi) + 2 - 2g \text{.}
\end{align*}
This is the usual Riemann--Hurwitz formula. 

We have shown to any pair consisting of a flat \(PSU(2)\) connection (equivalently, a projectively flat \(U(2)\) connection) and a holomorphic section of the corresponding \(\mathbb{C}P^1\) bundle gives rise to an Abelian exotic vortex equation on a constant curvature surface. 

Note that we could replace the target sphere with the flat plane or the hyperbolic plane. Correspondingly, one should change \(PSU(2)\) to \(E(2)\) or \(PSU(1,1)\) respectively. This allows one to find solutions with \(\kappa = 0\) or \(\kappa\) negative. Indeed, this was the original insight of Witten \cite{wittenMPS} for (nonexotic) hyperbolic vortices.

\subsubsection{Moduli}

It is conceivable that every solution to the relevant exotic vortex equation on a constant curvature surface arises from this construction. Showing this directly seems to be tricky as it involves integrating the equation \(\phi = \dd_a \psi\). Standard arguments as in \cite{mantonFVE}, for example, show that it is true locally. By studying the moduli problem for the pair \((a, \psi)\) we give some coarse evidence for the idea in more generality.

The first question is: what is the dimension of the moduli space of pairs of a flat \(PSU(2)\) connection \(a\) and a holomorphic section \(\psi\) of the associated \(\mathbb{C}P^1\)-bundle? The equations are
\begin{equation}
\label{eq:holeqs}
\begin{aligned}
F(a) &= 0 \\
\db_a \psi &= 0 \text{.}
\end{aligned}
\end{equation}

The linearisation of the first equation lives in the elliptic complex
\[
\Omega^{\bullet-1}( \mathfrak{su}(2))
\]
and so has real index
\[
6g-6 \text{.}
\]

On the other hand, the linearisation of the second equation lives in the complex
\[
\Omega^0(\psi^*T_{S/\Sigma}) \xrightarrow{\db_A} \Omega^1(\psi^*T_{S/\Sigma})
\]
and so has real index
\[
2 \text{deg}(\psi^*T_{S/\Sigma}) + (2-2g) \text{.}
\]
The vortex number is \(k = \text{deg}(\psi^*T_{S/\Sigma}) + 2g-2\), so this index is 
\[
2 k + 6 - 6g
\]

The combined system of equations \eqref{eq:holeqs} therefore has total index \(2k\), which is the same as the index of the vortex problem. This means that the `expected dimensions' of the two moduli problems agree. 

\subsubsection{The sphere, bad behaviour, and symmetry breaking}

The integrable Abelian vortex equation on the sphere is known as the Popov equation, after \cite{popovIV}. In the limit that the radius of the domain sphere becomes large, the constant \(\kappa_0 \to 0\), and the equation becomes the Jackiw--Pi equation of \cite{jackiwSSS}. 

The moduli space of pairs \((a, \psi)\) is directly accessible in this case. When \(\Sigma = S^2\), the only flat connection is the trivial one. The moduli space is then the space of rational functions \(\psi : \mathbb{C}P^1 \to \mathbb{C}P^1\), modulo the action of \(PSU(2)\) rotations of the target sphere.

This tells us that the moduli space of vortex number \(k\) solutions that can be constructed in this way is, as a manifold,
\[
\left( \mathbb{C}P^{k+3} - \Delta \right)/PSU(2)
\]
if \(k\) is even and empty if \(k\) is odd, where \(\Delta\) is the resultant hypersurface of rational functions that degenerate to one of lower degree.

The real dimension of this space is 
\[
2k + 3 \text{,}
\]
not the expected \(2k\). This isn't even even! If this is the full vortex moduli space, then it can't be symplectic, even though we expect it to be. What is going on?

The answer is that the moduli problem for flat \(PSU(2)\) connections on the sphere is not well-behaved. It has dimension 0, or dimension \(-3\) as a stack, rather than the expected dimension \(-6\) (which is its dimension as a derived stack).

One way to think about this is well illustrated in the zero vortex sector, which is now nontrivial. When \(k=0\) there are still solutions coming from rational functions \(\psi\) of degree 1. These form a group, the group \(PSL(2)\) of M\"obius transformations. Taking the quotient by \(PSU(2)\) gives 
\[
PSL(2)/PSU(2) \cong H^3 \text{,}
\]
hyperbolic 3-space. This is a kind of vacuum moduli space, which can be thought of as coming from the symmetry breaking \(PSL(2) \dashrightarrow PSU(2)\) implemented by the choice of round metric on the target sphere. 

This issue does not arise for nonexotic hyperbolic vortices, as in \cite{wittenMPS}, because all the holomorphic automorphisms of the hyperbolic plane (which is the target space) are isometries. The holomorphic symmetry is therefore unbroken by the metric and this strange \(k=0\) moduli space is not present.

A related and interesting point is raised in \cite{turnerQO}. There, Turner argues that the size modulus of a single Jackiw--Pi vortex should not be considered as a fluctuating mode in the quantum mechanical theory, because an infrared divergence renders it non-normalisable. Instead, Turner argues that the size modulus should be regarded as a fixed dimensionful parameter in the quantum theory. It seems plausible that this argument applies also to the other residual \(PSL(2)\) moduli and that, at least for the purposes of quantum mechanics, one should regard the moduli space as
\[
\left( \mathbb{C}P^{k+3}-\Delta \right)/PSL(2)\text{.}
\]
Pleasantly, this has complex dimension \(k\), which is the expected dimension of the moduli problem. The details surrounding the physical and mathematical implementation of this more severe quotient deserve further study.

There is another issue that we have not yet addressed: when \(\Sigma = S^2\), no vortex solutions with odd vortex number can be found with this construction. As discussed above, this is because there are no topologically nontrivial \(PSU(2)\)-bundles on the sphere and so the relevant moduli space of flat connections is empty.

\subsubsection{The Jackiw--Pi equation on the torus and Olesen's solution}

When \(\Sigma \) is a flat torus, the integrable exotic vortex equation is the Jackiw--Pi equation \cite{jackiwSSS}
\[
F(A) = \ii \kappa |\phi|^2 \text{.} 
\]
(Notice that this equation also arises as the infinite radius limit of the Popov equation considered previously. The difference here is in the boundary conditions.)

\begin{figure}[ht]
\centering
\begin{tikzpicture}
	\draw (0,0)--(4,0)--(4,4)--(0,4)--(0,0);
	\draw[dashed] (2,0)--(2,4);
	\draw[dashed] (0,2)--(4,2);
	\draw[fill=black] (1,1) circle (0.03);
	\draw[fill=black] (1,3) circle (0.03);
	\draw[fill=black] (3,1) circle (0.03);
	\draw[fill=black] (3,3) circle (0.03);
\end{tikzpicture}
\caption{A sketch of Olesen's construction of a single vortex solution to the Jackiw--Pi equation on the torus \cite{olesenSC}. One starts with a symmetric four-vortex solution, which can be produced from the elliptic function \eqref{eq:olesen}, and then cuts the domain into four, giving single vortex solutions on the quarter cells.}
\label{fig:olesen}
\end{figure}
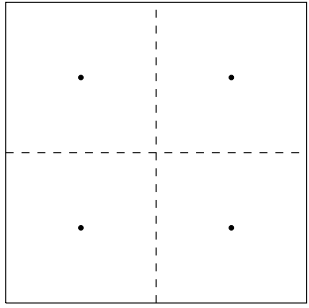

On the torus there exists a topologically nontrivial \(PSU(2)\)-bundle. This allows for the construction of solutions with odd vortex number, as in \cite{olesenSC}. As illustrated in \autoref{fig:olesen}, the idea of \cite{olesenSC} is to find a particular four-vortex solution on the torus (which can be built from an elliptic function) with the property that it defines a doubly-periodic single vortex solution on the quarter cell.

It is illustrative to consider the details of this. Let \(\Sigma\) be the torus, thought of as a rectangle with side lengths \(a\) and \(b\) in the upper half plane with opposite sides identified. Letting \(z\) be a complex coordinate on the upper-half plane, the symmetric elliptic function found by Olesen can be written 
\begin{equation}
\label{eq:olesen}
\psi_4 (z) = \frac{\wp(z) - e_3}{\sqrt{(e_3-e_1)(e_2-e_3)}}\text{,}
\end{equation}
where \(\wp\) is the Weierstrass elliptic function with periods \(a\) and \(b\) and
\begin{equation}
\label{eq:weierstrassconstants}
e_1 = \wp(a/2)\text{, } e_2 = \wp(b/2) \text{, } e_3 = \wp(-(a+b)/2) \text{.}
\end{equation}
Then the square of the Higgs field is
\[
\frac{|\psi_4'|^2}{(1+|\psi_4|^2)^2} \text{,}
\]
which has four zeroes and is periodic with periods \((a/2, b/2)\), not just \((a,b)\), as illustrated in \autoref{fig:olesen}.

What about \(\psi_4\)? As observed in \cite{olesenSC}, this is not periodic in \(a/2\) or \(b/2\). Instead, it varies by a M\"obius transformation as one moves by a half-period.
This means that it is a section of a \(\mathbb{C}P^1\)-bundle with nontrivial transition functions, which is the basic construction of this paper.

The general solution to the problem of solving the Jackiw--Pi equation on the torus was given in \cite{akerblomNCS}, generalising the ideas of Olesen. Indeed, the solutions all arise from `twisted' elliptic functions, referred to as \(\Omega\)-quasi-elliptic functions in \cite{akerblomNCS}. The observations that we have made about the topology of \(PSU(2)\) bundles resolves a question raised in \cite{akerblomNCS} about the topology-dependent nature of selection rules for Jackiw--Pi vortices.

\subsection{The nonAbelian case}
\label{subsec:exnonab}

Let \(X\) be a K\"ahler manifold with complex structure \(J\) and metric \(g\), and let \(\nabla\) be the corresponding Levi-Civita connection. Writing \(R\) for the curvature of \(\nabla\), we can then define a symmetric curvature tensor \(G\) by
\[
G(v,w) = R(v, J w) \in \Omega^0(\text{End}(TX))
\] 
for any pair of vectors \(v,w\). We will proceed as in the Abelian case, where \(X =\mathbb{C}P^1\), now using the tensor \(G\) to contract indices.

For now, let us not worry about the additional flat connection, and so work with a standard nonlinear sigma model. We consider holomorphic maps
\[
\psi : \Sigma \to X \text{.}
\]
Holomorphicity means that
\[
\dd \psi + J \circ \dd \psi \circ j_\Sigma = 0 \text{,}
\]
where \(j_\Sigma\) is the complex structure on \(\Sigma\).

Then \(\dd \psi\) is a holomorphic section of the Hermitian, holomorphic vector bundle \(V = K_\Sigma \otimes \psi^*T_X\). The Chern connection \(A\) on \(V\) is built from those on \(K_\Sigma\) and \(T_X\), giving
\[
F(A) = -F(\Sigma) \otimes \mathbb{1} + \psi^*F(\nabla)
\]
where \(\mathbb{1}\) is the identity endomorphism of \(\psi^*T_X\).

We now consider
\begin{equation}
\label{eq:curvquadratic}
(g_\Sigma^{-1})^{\mu\nu} G(\partial_\mu \psi, \partial_\nu \psi)\text{.}
\end{equation}
Using the holomorphicity of \(\psi\) and the definition of \(R\) in terms of \(F(\nabla)\), a Bogomolny-type rearrangement tells us that this is
\[
* \psi^* F(\nabla) \text{.}
\]

In analogy with the Abelian case, we would like to define
\[
\phi \stackrel{?}{\coloneqq} \dd \psi
\]
as a section of the holomorphic bundle \(V\). As the bundle \(V\) has rank \(n\), we expect this to be a Higgs field in a theory with \(N_c = n\) and \(N_f = 1\). 

Writing \(h\) for the Hermitian structure on \(V\), we have
\begin{align*}
(\phi \phi^{\dagger_h})^i_{j} &= \phi^i h_{jk} \bar{\phi}^k  \\
	&= (h_\Sigma^{-1})^{z\bar{z}} \partial_z \psi^i (h_X)_{jk} \partial_{\bar{z}} \bar{\psi}^k 	
\end{align*}
where \(h_\Sigma\) is the Hermitian structure on \(T_\Sigma\) and \(h_X = (h_X)_{ij} \dd \psi^i \dd \bar{\psi}^j\) is the Hermitian structure on \(\psi^*T_X\). By contracting the above expression with arbitrary vectors and expanding the Hermitian structure in terms of \(g\) and \(J\), we see that this is the same object as \eqref{eq:curvquadratic} up to a factor \(\kappa\) if
\begin{align*}
g(R(&u,v)w,x) = \\ 
&\frac{\kappa}{4} \left(g(u,w)g(v,x) - g(v,w)g(u,x) + g(u,Jw)g(v,Jx) - g(u,Jx)g(v,Jw) + 2g(u,Jv)g(w,Jx) \right)
\end{align*}
for vectors \(u,v,w,x\), which is equivalent to \(X\) having constant \emph{holomorphic sectional curvature} \(\kappa\) \cite{kobayashiFDG2}. The holomorphic sectional curvature is defined by
\[
H(\xi) \coloneqq g(R(\xi, \xi) \xi, \xi)/|\xi|^4 
\]
for \(\xi\) a holomorphic tangent vector.

Thus we find exotic vortex solutions from holomorphic curves in K\"ahler manifolds of constant holomorphic sectional curvature. In particular, the solutions that one finds are exotic vortex solutions in theories with \(N_c = \text{dim}_\mathbb{C}(X)\) colours and \(N_f = 1\) flavour. 

Manifolds with a complete K\"ahler metric of constant holomorphic sectional curvature are called \emph{complex space forms}, in analogy with the space forms of Riemannian geometry (which are complete Riemannian manifolds of constant sectional curvature). Every simply connected complex space form is either a complex projective space \(\mathbb{C}P^n\) (with \(H>0\)), a flat space \(\mathbb{C}^n\) (with \(H=0\)), or a hyperbolic space \(\mathbb{H}^n\) (with \(H<0\)) \cite{kobayashiFDG2}. This generalises the well-known uniformatisation theorem for \(n=1\).

In the context of exotic vortices, we are interested in the case of positive curvature, so we take \(X = \mathbb{C}P^n\).
Just as before, we can generalise this by introducing an additional flat connection for the isometry group of the target space \(X\). In the case that \(X = \mathbb{C}P^n\), the isometry group of \(X\) is \(PSU(n+1)\). 

Notice that this method also provides a way to generate solutions to the \emph{non-exotic} nonAbelian vortex equations, using holomorphic maps into the hyperbolic space \(\mathbb{H}^{N_c +1 }\) or quotients thereof. This generalises the result of \cite{wittenMPS}. Such maps do exist and it would be interesting to investigate the nature of the vortex solutions that can be generated in this way when \(N_c >1\).

\subsubsection{On moduli}

Fix a Riemann surface \(\Sigma\) and consider the theory of a flat \(PSU(n+1)\) connection on \(\Sigma\) and a section \(\psi\) of the associated \(\mathbb{C}P^n\)-bundle. 

The real index of the corresponding moduli problem is
\[
((n+1)^2-1)(2g-2) + 2 \psi^*c_1(T_{\mathbb{C}P^n}) [\Sigma] + n (2-2g) \text{.}
\]
The vortex number is the degree of the bundle \(V\) which is
\begin{equation}
\label{eq:RHNA}
k = n(2g-2) + \psi^*c_1(T_X) [\Sigma]  \text{.}
\end{equation}
In terms of \(k\), the real index of the moduli problem is
\begin{equation}
\label{eq:exoticind}
2k + n(1-n)(2-2g)\text{.}
\end{equation}

On the other hand, the real expected dimension of the moduli space of vortices in a \(U(N_c)\) gauge theory with \(N_f\) flavours is
\begin{equation}
\label{eq:vorind}
2k + N_c(N_f-N_c)(2-2g)\text{.}
\end{equation}
This is formally independent of the sign of the coupling constant, so it applies in the exotic case.

The solutions we find from twisted holomorphic curves are meant to be solutions in a theory with \(N_c = n\) and \(N_f = 1\). Comparing \eqref{eq:exoticind} and \eqref{eq:vorind}, we see that the expected dimensions of the moduli spaces of the relevant vortices and the relevant twisted holomorphic curves agree. This leads us to conjecture that all nonAbelian exotic vortex solutions with \(\kappa > 0\) on constant curvature backgrounds arise in this way.

\subsubsection{Selection rules}

In the Abelian case, we saw that integrable exotic vortices on the sphere must have even vortex number. This followed because of the non-existence of topologically nontrivial \(SO(3)\) bundles on the sphere and because of the Riemann--Hurwitz relation.

Here we have a similar story. Bundles with structure group \(PSU(n+1)\) are topologically classified by a \(\mathbb{Z}_{n+1}\)-valued invariant. More precisely, \(PSU(n+1)\) bundles on a Riemann surface \(\Sigma\) are classified topologically by \(H^1(\Sigma, \mathbb{Z}_{n+1})\). On the sphere, any \(PSU(n+1)\) bundle is necessarily trivial. 

For the trivial bundle, the Riemann--Hurwitz-type formula \eqref{eq:RHNA} tells us that the vortex number \(k\) must take the form
\[
k = n(2g-2) + (n+1)p
\]
for \(p \in \mathbb{Z}\). 
When nontrivial bundles exist, they allow for \(k - n(2g-2)\) to take nontrivial values modulo \(n+1\).

\subsection{Introducing curvature through uniformisation}

So far, our results require us to consider the vortex equations on constant curvature backgrounds. We can lift this requirement by introducing additional `uniformising data' in the solution. 

Let \(\Sigma\) be a Riemann surface with Hermitian metric \(h_\Sigma\). Any other such metric can be given as
\[
h_f = e^f h_\Sigma 
\]
for \(f\) some real function. 
The curvature of this metric is
\[
F(h_f) = F(h_\Sigma) +  \db\partial f \text{.}
\]
We would like to set \(F(h_f) = \ii \kappa_0 \omega_\Sigma\), where \(\kappa_0\) is a constant. This leads to a Poisson-type equation
\[
\db \partial f = \ii \kappa_0 \omega_\Sigma - F(h_\Sigma)
\]
for \(f\), which can be solved provided that \([\ii \kappa_0 \omega_\Sigma] = [F(h_\Sigma)]\) in \(H^2(\Sigma)\), which fixes \(\kappa_0\). The sign of \(\kappa_0\) is determined by topology through the Gauss--Bonnet theorem. To solve the equation we note that, if the condition above is satisfied, then \(\ii \kappa_0 \omega_\Sigma - F(h_\Sigma)\) is an exact form on \(\Sigma\) and we can use the \(\db \partial\)-lemma. Write \(f_0\) for a solution to this equation and write \(h_0 \coloneqq h_{f_0}\).

We now turn to the vortex equations. Previously, we found a solution \((V, \db_V, h ,\phi)\) in the holomorphic gauge by taking 
\[
V = K_\Sigma \otimes \psi^*T_\mc{X}
\]
for \(\psi\) a section of a certain \(X\)-bundle \(\mc{X}\), with \(X\) a K\"ahler manifold of constant holomorphic sectional curvature, and letting \(h\) be the natural tensor product metric \(h_\Sigma^{-1} \otimes h_{\psi^*T_\mc{X}}\).

When \(h_\Sigma\) does not have constant curvature, we simply let the Hermitian structure \(h\) be \(h_0^{-1} \otimes h_{\psi^*T_\mc{X}}\).  

While this saves us from constant curvature, it does not save us from global issues: the Fayet--Iliopoulos parameter in the vortex equation that this construction solves is still fixed by virtue of the Gauss--Bonnet theorem (although it can be varied by scaling the volume of \(\Sigma\)).

Putting everything together, we have shown the following. Let \(X\) be a complete K\"ahler manifold with constant holomorphic sectional curvature \(\kappa\) and isometry group \(H\). Then every pair consisting of a flat \(H\)-connection and a holomorphic section of the associated holomorphic \(X\)-bundle on a Riemannian surface \(\Sigma\) of genus \(g\) gives rise to a solution to the exotic vortex equations \eqref{eq:ev} in a \(U(\text{dim}_\mathbb{C}(X))\) gauge theory with one fundamental flavour with \(e^2 = -\kappa\) and \(\tau = \frac{2-2g}{2\pi \text{vol}(\Sigma) \kappa}\). Moreover, gauge equivalent pairs of this form give rise to gauge equivalent vortex solutions.

\section{The dynamics of exotic vortices}
\label{sec:dyn}

\subsection{Nonrelativistic Chern--Simons-matter theories}

The fact that the configuration space \(\mc{C}\) for the theory with \(e^2<0\) is naturally symplectic but not naturally K\"ahler implies that the dynamical \((2+1)\)-dimensional theory of exotic vortices should come from Hamiltonian dynamics on the configuration space. Thus the space \(\mc{C}\) is  the phase space of the \((2+1)\)-dimensional theory. 

This is exactly the setup of \cite{waltonVC}, where we studied nonrelativistic Chern--Simons-matter theories, realising them as theories of Hamiltonian mechanics on \(\mc{C}\). The story goes through \emph{mutatis mutandi} with \(e^2 <0\), so we do not repeat it here. The only subtle point concerns the sign of the Chern--Simons level. For our conventions, it turns out to be natural to accompany a flip in the sign of \(e^2\) with a flip in the sign of the Chern--Simons level.

The low-temperature theory is given by Hamiltonian mechanics on the moduli space of vortices with zero Hamiltonian, capturing the fact that the classical theory is dynamically trivial at low energy. As before, though, there is an interesting topological quantum mechanics at low temperature, obtained by geometric quantisation of the vortex moduli space \(\mc{M}\). 

Even though our understanding of the exotic vortex moduli space is limited, we can formally apply the ideas of \cite{waltonVC} to compute the Euler character of a (possibly virtual) quantum line bundle \(\mc{L} \to \mc{M}\) on the moduli space. Recall that such a bundle is one with first Chern class equal to the class of the symplectic form on \(\mc{M}\) divided by \(2\pi\). 

To properly carry out the geometric quantisation so as to obtain a space of states, one must choose a polarisation for the line bundle \(\mc{L}\) (corresponding, roughly, to a choice of `position' or `momentum' representation of the quantum states). In the non-exotic case, there is a natural way to do this, by giving \(\mc{L}\) the structure of a holomorphic line bundle. In the exotic case, \(\mc{M}\) is not necessarily a complex manifold in any natural way (or at all) so this is not a natural procedure. The quantity we compute is sufficiently `soft' that we need not worry about this. We will simply assume that a reasonable procedure exists and that the required index theorem still holds (to some extent, this follows formally from loop space supersymmetry methods). A more detailed study would have to deal with this potentially thorny issue.

This is not the only thorny issue in play. We saw that the moduli space of Popov-type vortices on the sphere was not even symplectic, even though it is `expected' to be. We must be aware of this kind of phenomenon. As discussed above, the arguments of \cite{turnerQO} suggest that one should omit the extra modes associated with non-rigid M\"obius transformations. If one does this, the moduli space does become even-dimensional and so could be symplectic in a natural way. The details of implementing this idea mathematically deserves study. Treating these badly behaved moduli spaces as symplectic manifolds in a formal sense presumably requires a kind of derived symplectic geometry. Physically, this could be understood by including `ghost' and `antighost' zero-modes in the moduli space (or, more-or-less equivalently, by working with the supersymmetric completion of the theory). We will not worry too much about this, proceeding in a rather formal fashion (the Euler character formally includes the ghost-like fields but does not allow us to disentangle the bosonic vortex moduli from the ghost moduli).

Such blindness to technical issues makes things rather easy: the general answers of \cite{waltonVC} do not (formally) require us to take \(e^2 > 0\).

\subsection{Exotic vortex counting}

For simplicity, we will illustrate the idea only for the Abelian case. It is shown in \cite{waltonVC} that the formal Euler character for \(k\) Abelian vortices with \(N_f\) flavours at Chern--Simons level \(\lambda\) on a Riemann surface of genus \(g\) is
\[
\sum_{j=0}^g \lambda^j N_f^{g-j} {g \choose j} {\lambda(\mc{A} - k ) + N_f(k-g+1) -1 \choose N_f(k - g+1) + (g-j) - 1}
\] 
where \(\mc{A} = \frac{e^2 \tau \text{vol}(\Sigma)}{4\pi}\).

It is natural to accompany a flip in the sign of \(e^2\) with a flip in the sign of \(\lambda\). We will see this explicitly in a moment. The fundamental physical reason for this is discussed in \cite{turnerQO}: it is related to the (in)stability of the theory under renormalisation group flow at the relevant fixed point. Thus, while in the non-exotic case we found that the theory was best-behaved when \(\lambda = N_f\), here we will consider the case of \(\lambda = -N_f\).

Indeed, the case of \(\lambda = -1\) provides a good window into the nature of the theory. When \(g=0\), one finds the index
\[
{ -\mc{A}  + 2k \choose k } \text{.}
\]
This captures the `Bradlow bound'
\[
-k < -\mc{A}
\]
for this theory correctly: the number of vortices \(k\) must be \emph{larger} that the normalised area \(\mc{A}\).

In particular, we see that Jackiw--Pi vortices (which have \(\mc{A} = 0\)) on the sphere at level \(-1\) have expected degeneracy
\begin{equation}
\label{eq:JPcount1}
{2k \choose k} \text{.}
\end{equation}
Of course, this should be supplemented with the knowledge that there are no vortex solutions when \(k\) is odd.
To some extent, this result gets to the heart of the question of why Jackiw--Pi vortices are interesting. When \(\tau = 0\), the theory is (at least na\"ively) gapless. However, this result indicates the potential existence of an interesting topological quantum mechanics at low temperature, which is usually a feature of gapped theories. 

The result \eqref{eq:JPcount1} is altered on surfaces of higher genus. For example, when \(g=1\), it is
\[
{2k -1 \choose k} - {2k-1 \choose k-1} \text{.}
\]
This is always zero! In fact, whenever \(g\) is odd, the terms in the index cancel pairwise, and the result is zero. This is probably because we are computing an index and not the true dimension of the Hilbert space, but it is intriguing nonetheless.

\section{Conclusions}
\label{sec:conc}

In this paper, we have addressed and generalised some of the questions raised in \cite{doroudSA} regarding the moduli and quantum dynamics of (exotic) vortices. We have shown that, just as the hyperbolic vortex equations are integrated by Blaschke products \cite{wittenMPS}, vortices in unitary gauge theories with a single flavour may be produced from twisted holomorphic maps into complex space forms. The holomorphic sectional curvature of the complex space form affects the type of vortex that one gets from this procedure. 
The construction sheds light on the strange flux quantisation rules seen for exotic vortices on \(S^2\). 

There are two particularly obvious avenues to extend the result. First, one could prove the converse statement - that this construction provides the general solution to the appropriate vortex equations. We have provided evidence for this result, the general proof of which is rendered somewhat tricky by nonAbelian gauge symmetry. Second, one could try to extend this construction to the case of multiple flavours.

We have also studied the classical and quantum dynamics of Abelian exotic vortices. We have formulated the general symplectic dynamics of exotic vortices and (formally) computed the corresponding Witten index of the quantum theory in the Abelian case. The results are intriguing, but strange. A more concrete, and less formal, approach (as in \cite{turnerQO}) may be needed for deeper insight.

\section*{Acknowledgements}

I am grateful to Nick Manton, Nick Dorey, and Bernd Schroers for helpful comments. This work has been partially supported by an EPSRC studentship and by STFC consolidated grants ST/P000681/1, ST/T000694/1.

\bibliographystyle{naturemag}
\bibliography{bibliography}

\end{document}